# Ptychographic non-line-of-sight imaging for depth-resolved visualization of hidden objects

Pengming Song[1,†], Qianhao Zhao[1,†,*], Ruihai Wang[1,†], Ninghe Liu[2], Yingqi Qiang[1], Tianbo Wang[1], Xincheng Zhang[1], Yi Zhang[1,3], and Guoan Zheng[1,4*]

[1]Department of Biomedical Engineering, University of Connecticut, Storrs 06269, USA
[2]Department of Electrical Engineering, California Institute of Technology, Pasadena 91125, USA
[3]Institute of Materials Science, University of Connecticut, Storrs 06269, USA
[4]Department of Electrical and Computer Engineering, University of Connecticut, Storrs 06269, USA
[†]These authors contributed equally to this work
[*]Email: qianhao.zhao@uconn.edu (Q. Z.) or guoan.zheng@uconn.edu (G. Z.)

**Abstract**: Non-line-of-sight (NLOS) imaging enables the visualization of objects hidden from direct view, with applications in surveillance, remote sensing, and light detection and ranging. Here, we introduce a NLOS imaging technique termed ptychographic NLOS (pNLOS), which leverages coded ptychography for depth-resolved imaging of obscured objects. Our approach involves scanning a laser spot on a wall to illuminate the hidden objects in an obscured region. The reflected wavefields from these objects then travel back to the wall, get modulated by the wall's complex-valued profile, and the resulting diffraction patterns are captured by a camera. By modulating the object wavefields, the wall surface serves the role of the coded layer as in coded ptychography. As we scan the laser spot to different positions, the reflected object wavefields on the wall translate accordingly, with the shifts varying for objects at different depths. This translational diversity enables the acquisition of a set of modulated diffraction patterns referred to as a ptychogram. By processing the ptychogram, we recover both the objects at different depths and the modulation profile of the wall surface. Experimental results demonstrate high-resolution, high-fidelity imaging of hidden objects, showcasing the potential of pNLOS for depth-aware vision beyond the direct line of sight.

The ability to image objects that are hidden from direct view has attracted significant interest in recent years[1, 2, 3]. Non-line-of-sight (NLOS) imaging exploits the light scattered off visible surfaces to recover information about the occluded objects or those located outside the direct line of sight. Various strategies have been developed in this field to infer object geometry and location[2, 4, 5, 6, 7, 8, 9, 10, 11, 12, 13, 14, 15, 16, 17, 18, 19, 20, 21]. For instance, time-of-flight techniques determine object characteristics by measuring the delay in light's return. This is typically done by illuminating a visible surface with short light pulses and measuring the temporal profile of the returned signals[2, 4, 5, 8, 11, 12, 13]. The collected transient measurements are then processed through algorithms to reconstruct the 3D shapes and reflectance properties of the objects[2, 4, 5, 8, 11]. Alternatively, speckle correlation-based techniques exploit the spatial correlations in scattered light, known as the memory effect[22, 23], to recover 2D projections and 3D structures of hidden scenes[14, 15, 16, 17]. In recent years, coherent illumination-based techniques have also shown promise in NLOS imaging. For example, by exploiting the coherence properties of scattered light at different wavelengths, it is possible to recover a holographic representation of the hidden object[18]. In addition to these strategies, other notable advances in the field have further expanded the capabilities of NLOS imaging[6, 7, 9, 13, 24, 25, 26, 27], bringing it closer to practical applications in autonomous navigation, search and rescue operations, and medical diagnostics.

Parallel to the development of NLOS imaging, ptychography has emerged as a powerful coherent diffraction imaging technique for both fundamental and applied sciences[28, 29, 30, 31, 32, 33, 34, 35, 36]. In a typical ptychographic setup, a spatially localized probe beam is used to illuminate the object in real space. The intensity of the resulting diffracted light waves is then acquired using an image sensor in reciprocal space. By translating the object to various lateral positions, ptychography acquires a series of diffraction patterns collectively referred to as a ptychogram. Each pattern within the ptychogram contains object information tied to the corresponding spatially confined regions in real space. The subsequent reconstruction process stitches these regions together to expand the effective imaging field of view while simultaneously recovering the complex-valued amplitude of the object's wavefield.

Despite its success in various applications, ptychography implemented with visible light often suffers from low imaging throughput and limited resolution[31]. To address these challenges, a recent innovation termed coded ptychography (CP) replaces the confined probe beam with an extended coded scattering surface placed directly on the image sensor[37, 38, 39]. This scattering surface serves as an effective 'illumination probe beam' as in conventional ptychography, modulating the complex-valued amplitude of the object's exit wavefields at a diffraction plane. By encoding the otherwise inaccessible high-resolution object information into intensity variations, CP enables super-resolution imaging beyond the limit imposed by the system's transfer function[37]. The operation is similar to that of structured illumination microscopy[40], where high-frequency object information is down-modulated into the passband of the optical system for detection. Notably, CP has demonstrated a numerical aperture (NA) of ~0.8 in the visible light regime, among the highest in lensless imaging demonstrations. Its imaging throughput is also comparable to or higher than the state-of-the-art whole slide scanners[37]. One particularly interesting development of CP is depth-resolved imaging[41]. By slight tilting the incident beam angle, CP can acquire a set of depth-multiplexed diffraction patterns. These patterns contain entangled information about objects at multiple depths, which can be reconstructed using a multi-slice ptychographic algorithm[42]. This depth-resolved capability of CP bears a strong resemblance to the depth-vision recovery sought in NLOS imaging.

The common thread between NLOS imaging and CP is the employment of a scattering surface in their configurations. In NLOS imaging, the surface is often a wall that scatters the light from the hidden object for detection outside the direct line of sight. In CP, the coded surface serves as an effective ptychographic probe beam for modulating the complex-valued object wavefields. Inspired by these similarities, we propose a new NLOS imaging approach termed ptychographic NLOS (pNLOS), which combines the principles of CP with NLOS imaging. In pNLOS, a focused laser spot is scanned across a wall to illuminate the hidden objects in the obscured region. The reflected wavefields from the hidden objects then travel back to the wall surface, get modulated by the wall's complex-valued profile, and the resulting diffraction patterns are captured by a regular camera. In pNLOS, the scanning laser spot on the wall serves as a virtual light source, illuminating the hidden objects with slightly different incident angles as in depth-multiplexed CP[41]. The wall surface in pNLOS modulates the object's complex-valued wavefields, serving the role of coded surface as in CP. As the laser spot is scanned to different positions, the objects' wavefields on the wall surface undergo corresponding translational shifts, with the amount of these shifts varying based on the depth of the objects. The translation of the objects' wavefields on the wall allows the acquisition of a depth-multiplexed ptychogram, an incoherent mixture of the diffraction patterns at different depths. By processing the ptychogram with multiplexed ptychographic algorithm[43, 44], we can recover object images at different depths as well as the unknown modulation profile of the wall surface. Built upon the foundations laid by both CP and NLOS imaging, the proposed pNLOS approach offers a fresh perspective on how these two domains can be combined to unlock new capabilities in imaging science.

## Results
### Principle and imaging model of pNLOS

Figure 1a illustrates the principle of pNLOS. In this setting, a focused laser spot is scanned across the scattering surface of the wall, serving as a virtual light source to illuminate the hidden objects from slightly different incident angles. The light waves reflected from the hidden objects then propagate back to the wall surface, and we denote the reflected object wavefield on the wall as $W_j(x, y)$, with the subscript $j$ denoting the $j^{th}$ object with a certain distance to the wall. In an example shown in Fig. 1a, there are 4 objects at different depths, with $j$ = 1, 2, 3, and 4. As the virtual light source translates to different lateral positions on the wall, the reflected object wavefield $W_j(x, y)$ also undergoes corresponding translational shifts $(x_{ji}, y_{ji})$ in Fig. 1a, where the subscript $j$ corresponds to the $j^{th}$ object and the subscript $i$ represents the $i^{th}$ positional shift during the light spot scanning process. The different colored lines in the right panel of Fig. 1a represent the trajectories of objects at different depths. Figure 1b shows the experimental setup, where the reflected wavefields from different objects are modulated by the complex profile of the wall surface, denoted as $CS(x, y)$, meaning coded surface. The resulting diffraction patterns, denoted as $I_i(x, y)$, are captured by a camera configured in a Scheimpflug arrangement (Fig. 1a). By adjusting the distance between the lens and the image sensor, we intentionally place the sensor at an out-of-focus position with a small defocus distance $d$, typically 0.5-1 mm. This defocus distance allows the conversion of the complex wavefield into intensity variations for detection, as demonstrated in both lens-based and lensless ptychography implementations[45, 46]. A similar arrangement can also be found in CP and the related synthetic aperture ptychography implementation, where a small defocus distance exists between the coded surface and the pixel array of the detector[37, 38, 39, 41, 47]. Conventional ptychography can be seen as a special case by removing the lens to capture the far-field diffraction pattern in reciprocal space[28].

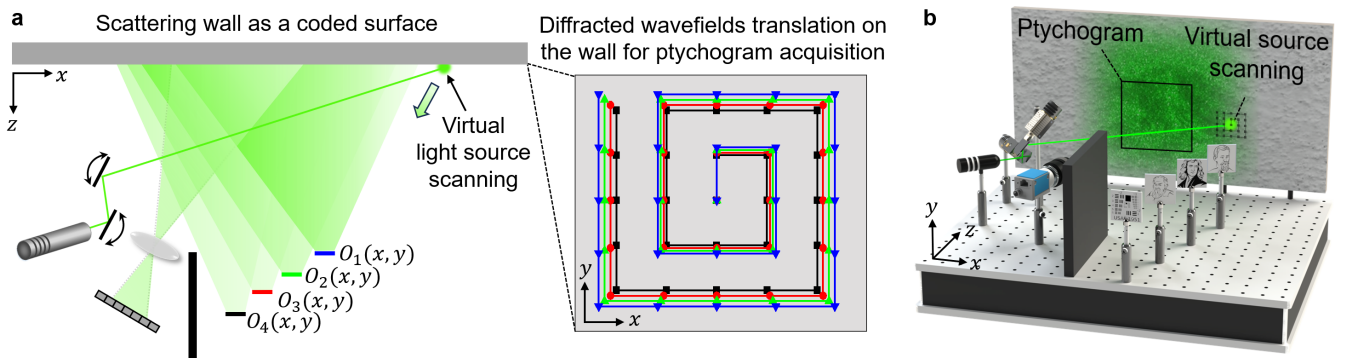

**Fig. 1| Schematics of ptychographic non-line-of-sight (pNLOS) imaging.** (a) In pNLOS, a scanning laser spot on the wall surface serves as a virtual light source to illuminate the hidden objects from slightly different incident angles. The scattering wall surface serves as a coded surface that modulates the complex-valued wavefields reflected from the hidden objects. As the laser spot is scanned to different positions, the resulting reflected wavefields on the wall surface also undergo translational shifts, with the amount of these shifts varying based on the depths of the objects. On the right panel, the color-coded lines represent the trajectories of objects at different depths. (b) The experimental setup, where a camera is used to capture the ptychogram on the wall. By computationally processing the ptychogram using a multiplexed reconstruction algorithm, pNLOS recovers images of the hidden objects at different depths as well as the unknown modulation profile of the scattering wall.

The forward imaging model for pNLOS can be expressed as:

$$I_i(x,y) = \sum_j \left| \{W_j(x - x_{ji}, y - y_{ji}) \cdot CS(x,y)\} * \text{PSF}_{free}(d) \right|^2, \tag{1}$$

where '·' indicates element-wise multiplication, and '∗' denotes convolution. The multiplication in Eq. (1) models the modulation process between the object exit wavefield on the wall surface and the complex modulation profile of the wall surface. The convolution kernel $psf_{free}(d)$ models free-space light propagation over a small distance $d$. It captures the diffraction effects that occur after the wavefield got modulated by the wall surface, similar to the free-space propagation between the coded surface and pixel array in CP. The diffraction pattern $I_i(x,y)$ is essentially an incoherent summation of the modulated diffraction patterns of different objects, as indicated by the summation over the subscript $j$ in Eq. (1). The subscript $i$ in $I_i(x,y)$ specifies the $i^{th}$ positional shift of the laser spot during the scanning process. By capturing a series of diffraction pattern $I_i(x,y)$ while scanning the laser spot on the wall, we obtain a depth-multiplexed ptychogram that contains entangled information about the objects at different depths. The subsequent reconstruction process aims to decouple this depth information, recovering both the object wavefields $W_j(x,y)$s and the unknown modulation profile of the wall surface $CS(x,y)$. With the recovered $W_j(x,y)$, one can then propagate the wavefield back to the correct object plane to recover the object profile $O_j(x,y)$. This operation is also similar to that of CP and synthetic aperture ptychography, which recover the exit wavefield at the plane of the coded surface and then refocuses it to the object plane post-measurement.

Directly recovering the depth-dependent object wavefield $W_j(x,y)$ on the wall surface present some challenges, primarily due to the absence of prior knowledge regarding the number of hidden objects and their positional shifts $(x_{ji}, y_{ji})$ during the laser spot scanning process. To address this issue, we perform an initial estimation of different objects by scaling the virtual light source's trajectory as positional shifts for objects at different depths. In this approach, different scaling factors correspond to object layers at different depths, similar to that in depth-multiplexed CP[41]. We then shift back the raw images according to the trajectories with different scale factors, generating a series of composite images. This shift-and-add process is analogous to back-projection in tomographic reconstruction[48]. To evaluate the quality of these composite images, we compute the Brenner index[49], a quantitative measure of image contrast, to assess the clarity and contrast enhancements of the back-shifted images. As shown in Supplementary Fig. S1, this analysis reveals peaks in the Brenner index plot, indicating the number of objects and their initial positional shifts $(x_{ji}, y_{ji})$ on the wall surface. With this initial estimation of the positional shifts, we further refine them via a cross-correlation analysis detailed in Methods and Supplementary Note 1.

With the recovered translational shifts, Supplementary Fig. S2 shows the detailed flowchart of the pNLOS recovery process. This process begins by generating initial estimates for the hidden objects by back-shifting the raw images according to their respective translational shifts. The wall's modulation profile is then initialized by averaging the captured images. Following this, the forward imaging model is applied to generate the complex wavefield $\psi_j(x,y)$ at the sensor plane, which then undergoes a magnitude projection step. The updated field $\psi'_j(x,y)$ at the sensor plane is subsequently propagated back to the wall surface to refine both the modulation profile $CS(x,y)$ and the object wavefield $W_j(x,y)$. The final step involves reconstructing each object $O_j(x,y)$ from its recovered wavefront $W_j(x,y)$. We approach it as an optimization problem that incorporates both model consistency and total variation denoising to improve the reconstruction quality. Supplementary Fig. S3 shows the simulation results of the pNLOS imaging process, demonstrating the effectiveness of the proposed method in reconstructing both the hidden object and the scattering surface.

**Experimental validation of pNLOS imaging**
To validate the capability of pNLOS imaging in recovering hidden objects, we first validate the method's effectiveness in reconstructing a single concealed object. In this experiment, a micro-circuit target was placed approximately 0.5 meters from the wall surface. The virtual light source was translated to a total of 1225 positions, covering a 3.5 mm × 3.5 mm area on the wall. Figure 2a displays the overall recovery result of the micro-circuit target using pNLOS imaging. To better illustrate the reconstruction quality, Figs. 2b1-d1 show the zoomed-in views of specific areas marked in Fig. 2a. The corresponding raw images (ptychogram) for these regions are presented in Figs. 2b2-d2, highlighting the speckle-like features under coherent illumination. For comparison, ground-truth images of the same regions captured using an upright microscope with a 4×, 0.2-NA objective lens are provided in Figure 2b3-d3. The agreement between the pNLOS reconstructions and the ground-truth images validates the effectiveness of the proposed method in reconstructing the structures of the hidden object. In addition to recovering the hidden object, pNLOS imaging also enables the reconstruction of the unknown modulation profile of the wall surface, which serves as the code surface (or the effective ptychographic probe beam) in the pNLOS imaging process. Figure 3 presents the recovered complex modulation profile of the wall surface in this experiment.

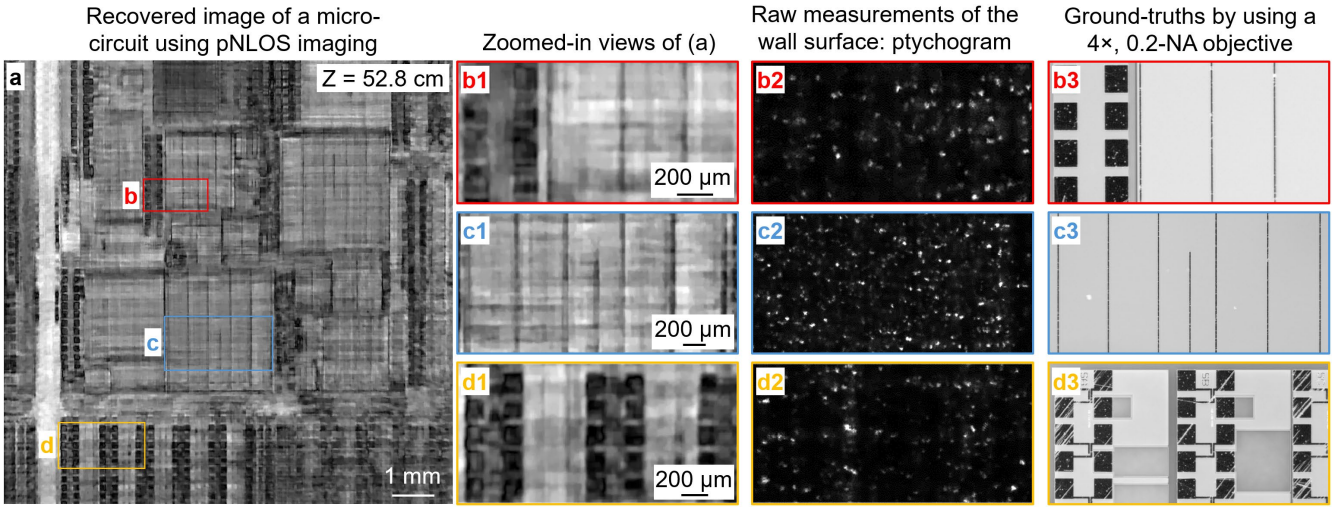

**Fig. 2| Validation of pNLOS imaging using a micro-circuit.** (a) The recovered image of the micro-circuit using pNLOS, revealing fine details of the features. (b1)-(d1) Zoomed-in views of (a). (b2)-(d2) The corresponding captured raw images with speckle-like features (the ptychogram). (b3)-(d3) The ground-truth images captured with a upright microscope with a 4×, 0.2-NA objective lens.

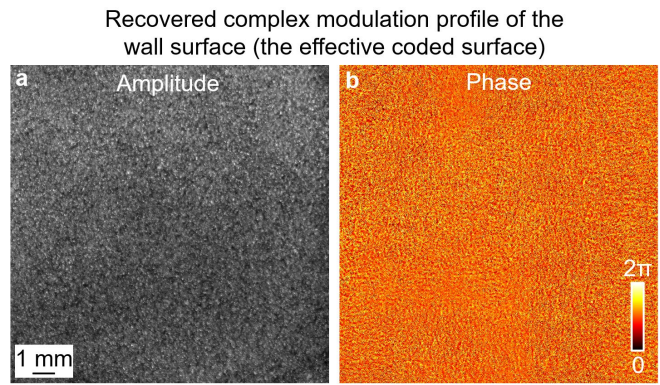

**Fig. 3| The recovered complex modulation profile of the wall surface.** (a) The recovered amplitude and (b) the phase.

To validate the depth-multiplexed imaging capability of pNLOS, we conducted another experiment using 4 objects placed at depths of 400 mm, 520 mm, 650 mm, and 1650 mm, respectively (Fig. 1c). These objects were slightly tilted so that their reflected wavefields overlap at the wall surface, creating an incoherent mixture of diffraction patterns for acquisition. Similar to the previous experiment, the laser spot was scanned across a 3.5 mm × 3.5 mm area on the wall, covering a total of 1225 positions (35 × 35 grid). The camera captured a single image at each laser position, resulting in a ptychogram containing 1225 frames. Figure 4a shows an example of a captured raw image, which contains an incoherent mixture of the diffracted wavefields from the objects at different depths. Figure 4b shows the ground-truth objects placed at different depths. Recovering the individual objects from this multiplexed measurement is a challenging task for conventional imaging methods. To tackle this problem, we applied the modified multiplexed ptychographic reconstruction algorithm[43, 44] in Fig. S2 to the captured ptychogram. This algorithm leverages the depth-dependent positional shifts of the object wavefields to demultiplex the captured intensity patterns and recover the objects at their respective depth locations. By iteratively updating the object wavefields and the coded surface profile in a manner that accounts for the multiplexing effect, the algorithm can effectively separate the wavefields from different depths and reconstruct high-quality images of the individual objects. Figure 4c shows the recovered objects using the pNLOS approach, demonstrating the ability of pNLOS to reconstruct multiple objects at different depths with minimum crosstalk between them. The reconstructed images also closely resemble their respective ground truth objects, showcasing the effectiveness of the depth-multiplexed pNLOS imaging approach.

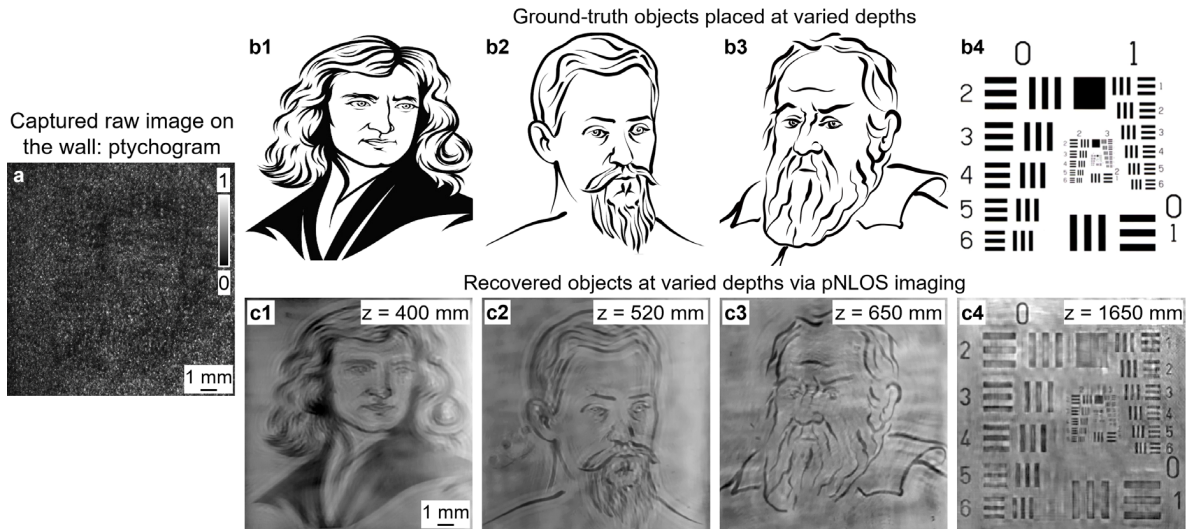

**Fig. 4| Validation of depth-multiplexed pNLOS imaging using 4 objects placed at varied depths.** (a) The captured raw image of the diffraction pattern on the wall surface, representing an incoherent mixture of diffracted wavefields at different depths. (b) The ground-truth objects placed at different depths. (c) The recovered objects using pNLOS imaging.

To quantitatively assess the resolution performance of pNLOS, we analyzed the reconstructed images of a USAF resolution target. In an experiment involving only the USAF target placed at a distance of 1650 mm from the wall (Supplementary Fig. S4a), pNLOS successfully resolved a linewidth of 56 μm, corresponding to group 3, element 3 of the target. For the depth-multiplexed case with 3 additional objects at different depths (Fig. 4c4 and Supplementary Fig. S4b), the achieved resolution experiences a degradation, with the system resolving a linewidth of 70 μm (group 2, element 6). To further investigate the depth resolution capabilities of pNLOS, we experimented with two objects placed at varying separations (Fig. S5a). The first object remained static at a distance of 400 mm from the wall, while the second object was incrementally moved to increase the gap ($\Delta z$) between them. Similar to the experiment in Fig. 4, we slightly tilted the second object so that its diffraction pattern overlapped with that of the first object on the wall surface. With this setting, we then recovered the image of the first object and analyzed the residual contrast of the second object in Fig. S5b. This residual contrast measures the unwanted crosstalk from the second object in the reconstruction of the first object. By setting the residual contrast to half of the initial residual contrast, the objects are considered sufficiently separated. Figures S6c1-c4 show the recovered images at different $\Delta z$ values (2 mm, 22 mm, 42 mm, and 52 mm). At small separations ($\Delta z$ = 2 mm and 22 mm), there is significant crosstalk between the two objects (Fig. S6c1 and c2). As the separation increases to 42 mm and 52 mm, the objects become clearly distinguishable, with minimal crosstalk (Fig. S6c3 and c4). By setting the residual contrast to half of the initial residual contrast, this analysis shows that the depth resolution is ~32 mm at a standoff distance of 400 mm.

To further demonstrate the depth-resolving capability of pNLOS, we experimented on imaging a tilted dragon-shaped object in Fig. 5. In this experiment, the object was placed at a distance ranging from 450 mm to 480 mm from the wall surface, with different parts of the object situated at different depths (Fig. 5a). This tilted geometry induces varying amounts of positional shifts in the reflected wavefields during the laser scanning process, providing a suitable test case for evaluating pNLOS's ability to recover depth information. To analyze the depth-dependent positional shifts, we divided the tilted object into 18 segments and estimated the shift of each segment from the captured ptychogram using the method described in Supplementary Note 1. The estimated positional shift vectors for the 9 sample segments are shown in Fig. 5b. Using the estimated positional shifts, we reconstructed each segment independently and calculated the Brenner gradient metric at different propagation depths ranging from 440 mm to 480 mm, with a 1 mm increment[49]. The resulting Brenner index curves for the 9 reconstructed segments are plotted in Fig. 5c. The peak of each curve indicates the depth position where the corresponding segment is in optimal focus. As expected, the peaks of the Brenner index curves occur at increasing depths for different segments, matching the tilted geometry of the object. In Fig. 5d, we show the recovered all-in-focus image of the entire object, generated by combining the in-focus reconstructions of all segments. This image is also color-coded based on their recovered depth positions, providing a visual representation of the object's 3D structure. The 3D rendering of the reconstructed object in Fig. 5e further illustrates the successful recovery of the tilted geometry, showcasing pNLOS's ability to capture depth information from the object's reflected wavefields on the wall surface. To assess the quality of the reconstructions, Figs. 5f1-i1 provide zoomed-in views of the recovered images while Figs. 5f2-i2 show the corresponding captured raw images.

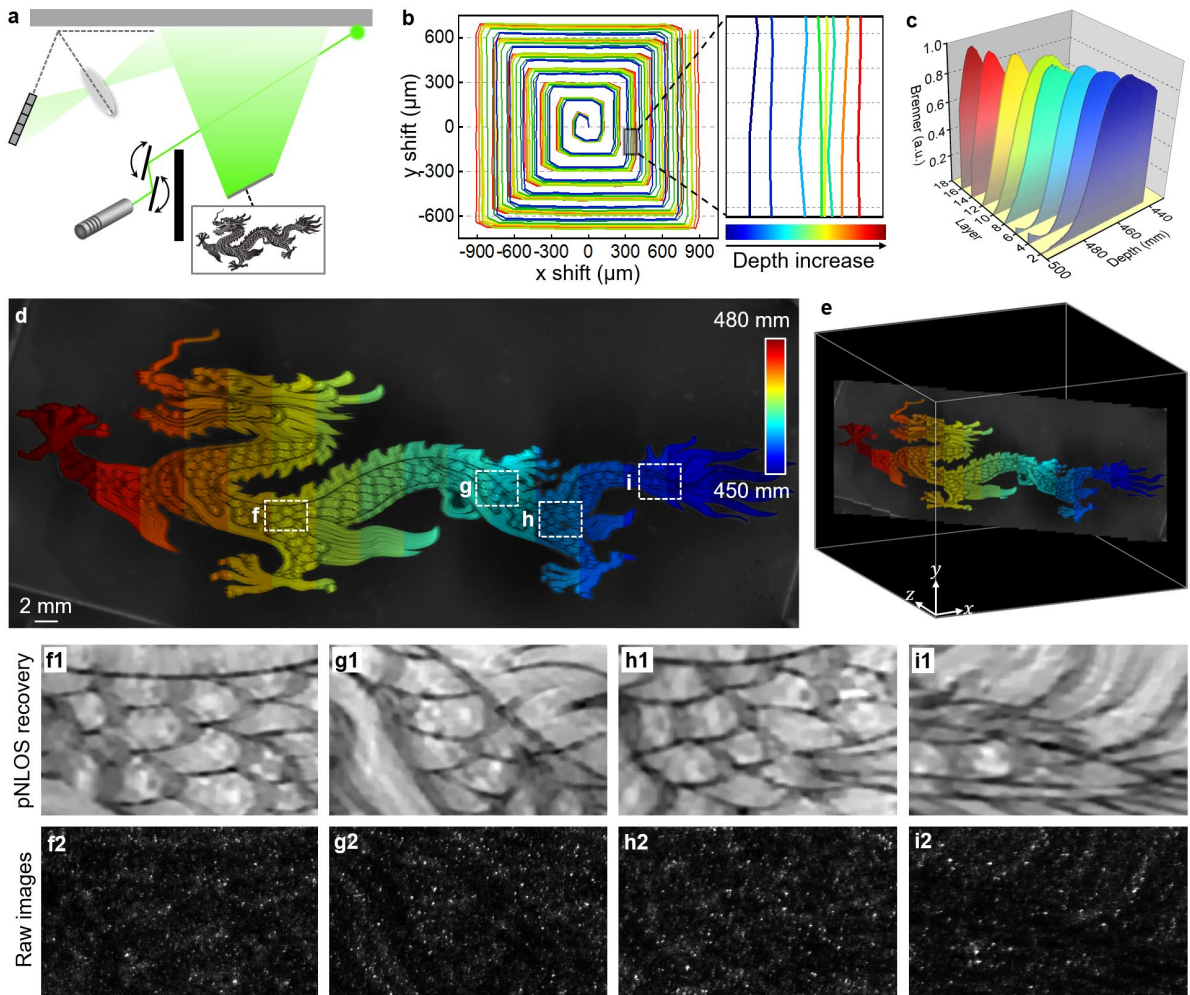

**Fig. 5| Reconstruction of a tilted sample.** (a) Schematic of the imaging setup with a tilted dragon sample, where different parts of the sample induce different positional shifts of the diffraction patterns on the wall surface. (b) The estimated positional shifts for 9 different sample segments of the tilted dragon. (c) Brenner index curves for inferring the best focus positions of different segments. These curves were obtained by propagating the recovered wavefields through a range of depths from 440 mm to 480 mm with 1 mm increment and calculating the corresponding Brenner index. The peaks of these curves indicate the optimal focus position for each segment of the tilted object. (d) The recovered all-in-focus image, color-coded by depth. (e) 3D rendering of the sample illustrating the focus positions of different segments. (f1-i1) Zoomed-in views of the recovered image in (d). (f2-i2) The corresponding captured raw images of (f1-i1).

Supplementary Fig. S6 offers additional insights into the depth-resolving capability of pNLOS imaging. Supplementary Fig. S6a shows the all-in-focused pNLOS recovery of the tilted dragon object, with a specific region selected for in-depth analysis. Supplementary Figs. S6b1-b3 present the reconstructions of the selected segment using three different positional shifts, corresponding to depths of 452 mm, 464 mm, and 480 mm. The best reconstruction quality of Supplementary Fig. S6b2 suggests that it most accurately aligns with the true depth of this particular segment. Supplementary Fig. S6c plots the Brenner gradient values for the three reconstructions. The peak of the curve corresponds to the reconstruction in Supplementary Fig. S6b2, confirming that it represents the most accurate depth for this part of the object. The successful reconstruction of the tilted object, with its gradually varying depth profile, showcases the potential of pNLOS for capturing depth information in complex hidden scenes.

**Discussion and conclusion**
The proposed pNLOS imaging technique introduces a new approach to recover depth-resolved information of hidden scenes by combining the principles of CP and coherent NLOS imaging. One of the main innovations of pNLOS is to treat the wall surface as a scattering coded layer to modulate the objects' diffracted wavefields. By scanning the laser spot across the wall surface, pNLOS imaging effectively translates the diffracted wavefields on the wall surface, introducing transverse translational shifts for acquiring a set of diffraction patterns termed a ptychogram. The Brenner index analysis in pNLOS also provides valuable insights into the depth distribution of hidden objects and their morphology. As demonstrated in the multi-object recovery experiment, the presence of multiple objects results in distinct peaks in the Brenner index curve, each corresponding to a different depth. This information is crucial for estimating the number of hidden objects and their respective positions. In the case of a single tilted object, the Brenner index exhibits a single peak that varies across different spatial positions of the captured image. These variations in the peak position reflect the changing depth of the tilted object, enabling the reconstruction of its 3D morphology. The depth-multiplexed ptychographic reconstruction in pNLOS is another innovation that enables the effective demultiplexing of the captured ptychogram and the recovery of individual objects at their respective depths. By leveraging the depth-dependent positional shifts and iteratively updating the object and wall surface estimates, the algorithm can handle mixtures of diffraction patterns and reconstruct multiple objects simultaneously, even when their wavefields overlap on the wall surface.

Drawing connections and distinctions between the proposed pNLOS and the related depth-multiplexed CP helps to clarify and summarize the advantages of our innovations. As shown in Supplementary Fig. S7, both pNLOS and depth-multiplexed CP employ a coded scattering layer to modulate the object wavefields and leverage the transverse translational diversity of the object wavefields to acquire a ptychogram. In pNLOS, we encode the depth information through the depth-dependent positional shifts of the object wavefields on the wall surface. Objects at different depths experience different amounts of shift during the laser scanning process. Similarly, depth-multiplexed CP encodes depth information through the angular diversity of the illumination beam, resulting in depth-dependent positional shifts in the captured diffraction patterns. However, there is a fundamental difference in how the two methods model the imaging process and reconstruct the object wavefields. Depth-multiplexed CP employs a sequential multi-slice model[42], where the object wavefield exits one layer, propagates to the next layer, gets modulated, and then propagates to the next layer, and so on. In the reconstruction process, depth-multiplexed CP sequentially updates different layers. In contrast, pNLOS captures the incoherent mixture of the diffracted wavefields from different layers, without explicit multi-slice modeling. To reconstruct the object wavefields, pNLOS employs a mixed-state ptychographic recovery algorithm, similar to that used in wavelength-multiplexed CP[43, 44]. This algorithm effectively reconstructs the individual object wavefields from the incoherent mixture of diffraction patterns, without the need for sequential layer-by-layer updates. Despite the difference, the successful demonstration of pNLOS highlights the potential of adapting and extending the principles of CP to the challenging domain of NLOS imaging. The innovations introduced in pNLOS, such as modeling the captured data as an incoherent mixture of ptychograms[43, 44] and developing a position-blind recovery algorithm, open up new possibilities for depth-resolved imaging in NLOS scenarios.

One current limitation of pNLOS is the low signal-to-noise ratio of the captured images, which necessitates a relatively long acquisition time (~1 second per image) and the use of samples with high reflectance. This limitation arises from the need to maintain a safe laser power level in our experiments to avoid damaging the wall surface. However, it is important to note that this is not a fundamental limitation of the pNLOS technique itself, but rather a practical constraint of the current experimental setup. Future improvements in the imaging system, such as the use full-field structured illumination strategies demonstrated in ptychography could potentially mitigate this issue[34, 45, 51, 52]. Additionally, the development of more efficient reconstruction algorithms under low light conditions[53] and the incorporation of prior knowledge about the scene could further enhance the SNR and reduce the acquisition time. Another challenge faced by the current pNLOS implementation is its susceptibility to ambient light. The presence of ambient illumination can interfere with the weak coherent signal scattered from the hidden objects, leading to a degradation in the reconstruction quality. To address this issue, future research could explore the use of time-gated detection or modulated illumination techniques to effectively suppress the ambient light and improve the SNR of the captured images.

To fully realize the potential of pNLOS, future research can focus on several areas. First, it is important to acknowledge that the depth resolution in pNLOS is fundamentally limited by factors such as the angular diversity of the laser illumination and the signal-to-noise ratio of the measured ptychogram. Future research efforts could focus on optimizing these factors to further improve the depth resolution and extend the imaging range of pNLOS. Secondly, adapting pNLOS to work with non-planar or rough wall surfaces could extend its versatility and broaden its range of applications. This can be potentially achieved by modelling the surfaces with effective spatially varying optical aberrations. Third, integrating pNLOS with polarimetric imaging techniques[54, 55] could provide additional information about the material properties and surface characteristics of the hidden objects, enhancing the overall interpretability. Finally, integrating pNLOS with other sensing modalities and machine learning techniques could enhance its reconstruction quality, speed, and robustness[56]. As research in NLOS imaging continues to advance, pNLOS offers new insights for depth-aware vision beyond the direct line of sight.

## Methods
### Experimental setup
The pNLOS imaging system consists of a continuous-wave laser (Laser Quantum GEM 532 nm wavelength), a galvanometer mirror system for laser scanning (Dual Axis ScannerMAX), a camera with a 20-megapixel image sensor (Sony IMX 183), and a scattering wall surface. The camera's sensor plane was aligned with respect to the scattering wall surface using a Scheimpflug arrangement, which ensures that the entire field of view is in focus despite the tilted object plane. The laser is focused onto the wall surface, creating a spot size of approximately 500 µm in diameter. The galvanometer mirror system scans the laser spot across a predefined area on the wall, typically covering a 3.5 mm × 3.5 mm grid with 1225 scanning points (35 × 35). We aligned the galvanometer mirror system to ensure that the laser spot scanned the desired area on the wall surface with high precision. The captured images are then processed using the pNLOS reconstruction algorithm to recover the depth-resolved images of the hidden scene.

### Blind recovery of the positional shifts
Unlike conventional NLOS imaging with temporal measurements, one of the key challenges in pNLOS imaging is the blind recovery of the positional shifts induced by objects at different depths. Since the depth and position of the hidden objects are unknown a priori, the positional shifts cannot be directly measured or calculated. Instead, they must be estimated from the captured ptychogram itself. To address this challenge, we developed a two-step approach for estimating the positional shifts, as detailed in Supplementary Note 1. First, we use cross-correlation analysis and a tomography-based shift-and-add technique[48, 57] combined with the Brenner gradient[49] to identify the hidden objects, determine their respective scale parameters, and estimate the initial positional shifts. This process also shares some similarities with the process on identifying different layers in depth-multiplexed CP[41]. Next, these initial estimates are refined through an iterative process involving the generation of reference images for each layer and the updating of positional shifts based on cross-correlation between the updated reference images and the captured images[58]. The successful implementation of this method, as demonstrated in Fig. 4 and Supplementary Fig. S1, highlights the effectiveness of the pNLOS imaging system in handling complex, depth-multiplexed scenarios without prior knowledge of the object positions or the scattering surface profile.

**Depth-multiplexed ptychographic reconstruction**
With the recovered translational shifts, the pNLOS reconstruction algorithm aims to demultiplex the captured ptychogram and recover the individual object wavefields from the incoherently mixed measurements. Supplementary Fig. S2 shows the detailed flowchart of the pNLOS recovery process, which operates in an iterative manner, alternating between updates in the object domain and the measurement domain. The process begins by generating initial estimates for the hidden objects by back-shifting the raw images according to their respective translational shifts obtained from the blind shift recovery step. The wall's modulation profile, which represents the unknown scattering properties of the wall surface, is initialized by averaging the captured images. Next, the forward imaging model is applied to generate the complex wavefield at the sensor plane. This wavefield undergoes a magnitude projection step[29], where the phase information is retained but the magnitude is replaced with the square root of the measured intensity from the ptychogram. The updated field at the sensor plane is then propagated back to the wall surface using the angular spectrum method[59]. The back-propagated wavefield is used to refine both the modulation profile of the wall and the object wavefield at the wall surface. This refinement process is carried out using a mixed-state ptychographic recovery algorithm, similar to that employed in wavelength-multiplexed CP[35, 43, 44]. The final step of the reconstruction process involves recovering the object profile from its corresponding wavefield at the wall surface.

Supplementary Fig. S3 shows the simulation results of the pNLOS imaging process. The ground truth of the object is shown in Supplementary Fig. S3a1-a2, while the ground truth of the scattering wall surface is presented in Fig. S3b1-b2. The captured raw images, which are the input to the pNLOS reconstruction algorithm, are displayed in Supplementary Fig. S3c1-c3. These raw images exhibit a complex mixture of the object's diffraction patterns modulated by the scattering surface. The recovered object using the pNLOS imaging approach is shown in Supplementary Fig. S3d1-d2, closely resembling the ground truth in both intensity and phase. Similarly, the recovered scattering surface (Supplementary Fig. S3e1-e2) reconstructs the complex modulation profile of the scattering medium. These simulation results validate the capability of pNLOS in simultaneously recovering the hidden object and the unknown scattering surface from the captured raw images.

**Acknowledgements**
This work was partially supported by the UConn SPARK Grant (G. Z.), the UConn Faculty Scholar Award (G. Z.), and the National Science Foundation 2012140 (G. Z.). Q. Z. acknowledges the support of the General Electric fellowship. P. S. acknowledges the support of the Thermo Fisher Scientific fellowship.

**Author contributions**
G. Z. conceived the concept of pNLOS and supervised the project. P. S. and Q. Z. developed the prototype systems. P. S., Q. Z., and R. W. performed image reconstruction. P. S., Q. Z. and R. W. prepared the display items. Y. Q., T. W., X. Z., and Y. Z. prepared the samples for imaging and analysed the data. All authors contributed to the discussing, writing, and revision of the manuscript.

**Disclosures**
The authors declare no conflicts of interest.

**Data and code availability**
Relevant data and code can be obtained from corresponding authors upon reasonable request.